\documentclass[12pt,preprint]{aastex}
\usepackage{natbib}
\usepackage{graphicx}
\slugcomment{draft version}
\usepackage{ulem}
\begin{document}
\title{Reverse-Shock in Tycho's Supernova Remnant}

\author{F.J. Lu\altaffilmark{1}, M. Y. Ge\altaffilmark{1}, S. J. Zheng\altaffilmark{1},
S. N. Zhang\altaffilmark{1}, X. Long\altaffilmark{1}, and B. Aschenbach\altaffilmark{2}}

\affil{$^1$Key Laboratory for Particle Astrophysics, Institute of High Energy
Physics, Chinese Academy of Sciences, Beijing 100049, P.R. China;
lufj@ihep.ac.cn; gemy@ihep.ac.cn}
\affil{$^2$PR Vaterstetten, Mozartstra$\ss$e 8, 85591 Vaterstetten, Germany}

\begin{abstract}
Thermal X-ray emission from young supernova remnants (SNRs) is
 usually dominated by the emission lines of the supernova (SN) ejecta,
which are widely believed being crossed and thus heated by the inwards propagating
reverse shock (RS).
Previous works using  imaging X-ray data have shown that the ejecta are heated
by the RS by locating the peak emission region of the most recently ionized matter,
which is found well separated towards the inside from the outermost boundary.
Here we report the discovery of a systematic
increase of the Sulfur (S) to Silicon (Si) K$\alpha$ line flux ratio with
radius in Tycho's SNR. This allows us, for the first time, to present continuous
radial profiles of the ionization age and, furthermore, the elapsed ionization
time since the onset of the ionization, which tells the propagation history
of the ionization front into the SNR ejecta.
\end{abstract}


\section{Introduction}

The reverse shock (RS) plays a key role in the study of explosive phenomena such as supernova (SNe) and
gamma-ray bursts (GRBs). When a SN
explodes, a blast wave (BW) is generated and propagates into the ambient medium, with
fast moving ejecta following behind. In the 1970s, it was proposed that an
inward-propagating shock wave, the so-called RS, will accompany the deceleration
of the ejecta by the ambient medium, and the ejecta behind this shock can be a
strong source of thermal X-rays (Gull 1973; Rosenberg \& Scheuer 1973; McKee 1974; Gull 1975).
As the ejecta expand, their density and pressure decrease, which causes the
RS to accelerate (McKee 1974), and simulations show that in supernova remnants (SNRs) the velocity of
the RS could be eventually much higher than that of the BW (Truelove \& McKee 1999),
resulting in a hot core enclosed by a cooler shell (Vink 2012).
The X-ray emission from the RS-heated ejecta thus provides important information
about the chemical evolution of stars and galaxies. In GRB study, the RS
is used to explain the early time flares and emission excess in the optical and
radio bands and is regarded as the most useful probe of the initial bulk Lorentz
factor and the ejecta composition (M\'eszaros \& Rees 1993; M\'eszaros 2002;
Sari \& Piran 1999; Kobayashi \& Zhang 2003; Laskar et al. 2013).
The RS propagating back into the accretion flow from the normal companion star is
also suggested to be the cause of a weak Hydrogen emission line that varies in radial velocity in a
cataclysmic variable system (Krauland et al. 2013). Therefore, the RS is important for a wide range
of astrophysics study.

The best sites to study the dynamics of the RS are SNRs because of their
fairly large angular sizes. Observations and analyses of the X-ray emission
from a few young SNRs have established the action of RS, as the X-ray
emission properties of plasma are
related to its shock heating history. After matter is crossed and heated by a
shock, the atoms start to be ionized gradually, and the ionization
age ($\tau=n_{e}t$, where $n_e$ is the electron number density and $t$ the
time span since the shock heating) is used in SNR study to describe the ionization
state of plasma (Itoh 1977). Since the line emission properties of the ejecta in a SNR
change with $\tau$, spatial variation of those properties could be used to trace the
propagation of the shock that heats the ejecta. In the RS scenario, the ejecta in
the outer layers should generally have a higher $\tau$ than in the inner layers,
and evolution of line emission properties with radius is thus expected. Most of the
evidence for RS is obtained by this method so far.

X-ray imaging spectroscopy of the Tycho's SNR by Hwang and Gotthelf (1997) using
the {\it ASCA} data showed that for the Tycho's SNR
the Fe K$\alpha$ line emission has its peak at a smaller
radius than the Fe-L and Si-K emission. Based on a detailed broad band X-ray
spectral analysis of the {\it ASCA} data
Hwang et al. (1998) determined a significantly lower $\tau$  for the Fe-K band emission
than for the Fe-L and Si-K emission bands, which they attributed to a more
recent heating of the  Fe-K line emitting ejecta by the RS.
Warren et al. (2005) used the so-called Principal
Component Analysis technique to study images and spectra taken with the Chandra X-ray
Observatory ({hereafter \it Chandra})
to disentangle between emission regions which are either
line-dominated or featureless-dominated.
On the basis of the obtained morphology the authors determined the locations of the BW, the contact
discontinuity and the inner edge of the RS, such that $r=0.7$ with $r$ the ratio of the radii of the
RS over that of the BW.
They also suggest that the Fe-K band emission comes from the innermost portion of the shocked
ejecta and its inner edge denotes the location of the RS.
These interpretations would require some stratification of the elements throughout the explosion and
subsequently, at least as far as  the bulk of the Fe emission is concerned, which has to remain
inside of the region populated by the
intermediate-mass elements like Si and S. To be consistent with the high temperatures required for the
Fe-K band emission observed by Hwang and Gotthelf (1997), an additional heating process of the electrons
in the area of the RS is needed, e.g., collisionless electron heating, as proposed by Badenes
et al. (2005). One-dimensional delayed detonation models for
the SN coupled with non-equilibrium ionization of the shock
heated ambient plasma and the RS heated ejecta were compared
with the images and spatially integrated spectra of this remnant
obtained with {\it Chandra}
and {\it XMM-Newton}, which could be well reproduced with some
degree of chemical stratification of the ejecta and the Fe-K
emission peaking interior to the intermediate-mass element distribution (Badenes et al. 2006).

Application of the same kind of approach leads to a satisfactory
agreement with the observations of SNR 0509-67.5 located in the Large
Magellanic Cloud (Badenes et al. 2008).
Similarly, the {\it Chandra} X-ray observations of Kepler's SNR could be explained
with hydrodynamical modeling by Patnaude et al. 2012.
 For SN1006 the {\it Suzaku} measurements are consistent with chemical
stratification but based on the
spatially integrated X-ray spectra the RS may not have reached or
heated the supposedly interior Fe shell (Yamaguchi et al. 2008).
The topic of chemical stratification of the elements in young SNR,
which is highly relevant to the supposedly innermost Fe-K band
emission and therefore to possibly the innermost
edge of the RS is further addressed by Yamaguchi et al. (2014a)
for Tycho's SNR, and Yamaguchi et al. (2014b),
more generally, for young and middle-aged SNRs with Fe K$\alpha$ emission.
Yamaguchi et al. (2014a) report the detection of Fe K$\beta$ fluorescence
emission from low ionization Fe at a radius even smaller
than the peak radius of the Fe$\alpha$ line emission at $r=0.66$.

In addition to the above facts obtained from the line emission properties,
there are some other evidence supporting the RS scenario. In Cassiopeia A, a
sharp rise of the radio and Si emissivity at the inner edge of the bright
X-ray ring was suggested to be associated with the RS (Hwang et al. 2000).
It has also been shown
that the available multi-wavelength observations in the radio, X-ray,
and gamma-ray bands of Cassiopeia A can be best explained by invoking particle
acceleration by both forward and RSs (Zirakashvili et al. 2014).
In SN1987 A, the broad Ly$\alpha$ emission feature located inside the
inner ring was attributed to the excitation by a RS (Sonneborn et al. 1998;
Michael et al. 2003),
which would indicate the position of the RS front. Hubble Space
Telescope observations of a star located behind the
SN 1006 remnant revealed that the ejecta had been decelerated probably by the
RS by $44\pm11$\,km\,s$^{-1}$ from year 1999 to 2010 (Winkler et al. 2011).

Summarizing the previous works, the observations and subsequent
analyses have allowed to locate the bulk or peaks of the emission of
the most recently shocked matter, by which the existence of the RS
in these SNRs has been demonstrated. In this paper we would like to
go one step further, i.e., outlining the evolution of the RS by
measuring the radial profile of the ionization age from the outer
edge of the ejecta, or the so called contact discontinuity,
as far as possible towards the center of Tycho's SNR by
using the data from the archived long exposure observations made
in 2009 with {\it Chandra}.
Tycho's SNR is relatively nearby ($\sim3$\,kpc) (de Vaucouleurs 1985;
Hern\'andez et al. 2009), bright in X-rays, and has a quite symmetric
morphology (Lu et al. 2011), which means that the shock structure
should be regular and thus easy to be revealed, and it is indeed
the most widely studied SNR as to the existence of RS. We also
attempt to separate electron density and the time since the onset
of the RS shock heating at a given radius, i.e., a time profile for
the ionization. In this way the progression of the RS towards the
interior of the SNR is documented rather than determining just the
location of its inner edge.

Instead of doing X-ray spectral modeling to get the $\tau$ of the
plasma in different regions, here we have searched for signals of the
RS kinematics by using the relative strength of the silicon (Si)
and sulfur (S) K$\alpha$ emission lines across the Tycho's SNR.
With the high spatial and spectral resolution X-ray observations
of young SNRs available, one may think that it is possible to obtain
$\tau$ of the plasma in a region by modelling its X-ray spectrum, and
the RS behavior could be revealed with such analyses for a series
of spectra taken at different radii. However, it is very difficult
in practice, because the plasma along the line of sight usually
contains many clumps that have quite different abundances,
temperatures, densities and ionization states. In contrast, using
the relative strength of the Si and S lines is more robust and
straight forward.
In various SN models, Si and S are synthesized in similar physical
conditions and are therefore expected to be well mixed spatially,
or with S locating only very slightly inside Si, at most
(Nomoto 1997; Iwamoto et al. 1999; H\"oflich et al. 2002;
H\"oflich et al. 2006). This is also confirmed observationally for
Tycho's SNR by the similar overall temperature and  $\tau$ of these two
elements (Hwang et al. 1998). Therefore, Si and S should be very similar
in temperature, density, and $\tau$ distributions, and the comparison
of their line emission strengths could be used to trace the
propagation of the RS.

\section{Observation and Results}

The data we analysed were obtained in 2009 with the imaging array of the Advanced CCD
Imaging Spectrometer (ACIS-I) aboard {\it Chandra}
for a total effective exposure of 734\,ks. The observations are listed in Table 1.
The data were calibrated with Chandra Interactive Analysis of Observations
(CIAO V4.5) following the standard procedure\footnote{http://cxc.harvard.edu/ciao}.
Throughout this paper the uncertainties are at $1\sigma$ level.

\subsection{X-ray Images}
Fig. 1 presents the X-ray maps of the remnant in 0.5-8.0, 1.6-2.0, and 2.2-2.6\,keV energy bands,
as well as its 2.2-2.6\,keV to 1.6-2.0\,keV intensity ratio map. Intensities in 2.2-2.6\,keV
and 1.6-2.0\,keV are dominated by the Helium-like (He-like) S and Si K$\alpha$ lines, respectively.
Therefore, Fig. 1(d) clearly shows that the S/Si line intensity ratio in an outer shell is higher
than that in the inner part.

\subsection{Radial distributions of the Si and S line fluxes and their ratio}

To study the variations of the line flux ratio at different radii
in a more quantitative way, spectra of three series of annuli
have been extracted and fitted to get the fluxes of the Si K$\alpha$($\sim$1.86\,keV) and
S K$\alpha$ ($\sim$2.45\,keV) lines. Regions from which the annulus series were extracted are denoted
in Fig. 1.  The width of the annuli
is 4$^{\prime\prime}$ in the outer regions, increases to 8$^{\prime\prime}$ in the inner regions,
and becomes even wider in the inmost fans, so as to have enough photons for each spectrum.
The background spectrum was extracted from regions outside of the remnant, and the spectra were
analyzed with XSPEC (v12.7.1).The spectrum of an
annulus was fitted with a two-component VNEI model (Borkowski et al. 2001)
 that describes the X-ray emission spectrum of plasma in non-equilibrium ionization state,
with the temperatures, emission measures, ionization age $\tau$, abundances of the
elements as free parameters and the redshift fixed at zero. The WABS model
(Morrison \& McCammon 1983) with the column density fixed at $6\times10^{21}$\,cm$^{-2}$ was
used to account for the interstellar
medium absorption, and a Gaussian line at 3.1\,keV was added to represent the Ar emission lines,
which is missing in the VNEI model. The absorption column density used here is similar to that
of Warren et al. (2005), and to set it at $5\times10^{21}$\,cm$^{-2}$ or $7\times10^{21}$\,cm$^{-2}$
does not change our main results significantly.

Now we describe the process to get the fluxes of the lines. For Si, we first
fitted the spectrum with a two-component VNEI model, then set the Si
abundance to zero, and used
four Gaussian components to account for the emission from Si.
In the four Gaussian components, one with the central
energy at about 1.85\,keV corresponds to the prominent K$\alpha$ line,
the other three with central energies fixed at 2.006, 2.182 and 2.294
keV and width fixed as 10$^{-5}$ keV represent the Ly$\alpha$,  the He$\beta$,
and the He$\gamma$ lines, respectively. The He$\gamma$ flux was set as 0.55
He$\beta$ flux in the fitting process, which is typical for a plasma around
1 keV as in this remnant (e.g., Hwang \& Gotthelf 1997).
The flux of the Gaussian component at $\sim$1.85 keV is
used in our S/Si line flux ratio study. For S, the analysis
process is the same.  The flux of the first component at about 2.45\,keV
represents the He K$\alpha$ line and is used for
the S/Si line flux ratio study, while the other three components
with central energies fixed at 2.623, 2.884 and 3.033 keV and
width fixed as 10$^{-5}$ keV represent the Ly$\alpha$,  the He$\beta$ and the
He$\gamma$ lines of S, respectively. The S He$\gamma$ flux was set as 0.56
He$\beta$ flux, also following Hwang \& Gotthelf (1997).
 Fig. 2 illustrates an example of the spectral fitting
results, and Table 2 lists the fitted parameters of the
corresponding Gaussian components.
The S to Si line flux ratios versus distances from the
geometrical center (Warren et al. 2005)
for these regions are plotted in Fig. 3. Apparently, the ratio increases with
radius for every of the three regions,
except of some small bumps caused by bright fragments
in projection.

The errors of the line fluxes plotted in Fig. 2 and those listed in Table 2
are obtained in the spectral fitting process by using the {\it error} command
in XSPEC. In this process the continuum emission was fixed,
and so the line flux uncertainty induced by the continuum determination
had not been included. We thus used
a Monte-Carlo method to estimate how significant this fraction of error is.
For  the spectrum shown in Fig 2, as an
example, we sampled it 500 times and got 500 simulated spectra. Then we
repeated the spectral fitting and Gaussian
fitting processes and got 500 pairs of Si and S K$\alpha$ line flux values. The distributions of the Si and S
line fluxes as well as their ratios are presented in Fig 4, whose widths could
be taken as the statistical errors of these quantities. Fitting the distributions with a
Gaussian function gives the $1\sigma$ values of 1.5\% and 3.7\% for the Si and S lines,
respectively, quite similar to those listed in Table 2, which are 1.4\% and 3.2\%.
 Therefore the contribution of the continuum determination to the line flux uncertainty
could be neglected, and the line flux uncertainties used in this paper are thus directly
derived by the  {\it error} command of XSPEC.

\section{Discussion}

The observed radial distributions of the S/Si line flux ratios provide an opportunity to
study the shock heating history of the plasma. As introduced earlier, S distributes
very similarly to or at most a little closer to the inside than Si, the S/Si
abundance ratio should be almost
constant or slightly decrease with radius (e.g., Nomoto 1997; Iwamoto et al. 1999; H\"oflich et al. 2002;
H\"oflich et al. 2006).
The observed S/Si line flux ratios that increase with radius are thus not the results
of abundance changes, but the variation
of $\tau$ and/or temperature of the plasma versus radius, which will be discussed below.

In Fig. 5, the S/Si flux ratios as a function of $\tau$ (predicted by the VNEI model)
are plotted for various electron temperatures, where the S/Si abundance ratio is assumed
to be 1.35 times the solar value (Hwang \& Gotthelf 1997).
Apparently, a high S/Si K$\alpha$ line flux ratio corresponds to a higher temperature and/or $\tau$,
and thus the observed S to Si line flux ratio profiles plotted in Fig. 3 are the consequence
of increasing temperature and/or $\tau$ with radius. However, previous observations of the Tycho's SNR
found that the temperature is actually higher in the interior (Decourchelle et al. 2001;
Warren et al. 2005), which is also supported by numerical simulations (Badenes et al. 2006).
Thus the only explanation of Fig. 3 is that $\tau$ increases with radius.

In order to further find out which of $n_{e}$ and $t$ dominates the observed $\tau$ variations,
we deprojected the observed radial profiles of Si, S, and thermal continuum fluxes to obtain
the S/Si line flux ratios and the scaled emission measure in real space. The
observed surface flux $F(r)$ in a direction is the integration
of the volume flux $F_{d}(r)$ along the line of sight. Under the assumption of
spherical symmetry the formula is (Helder \& Vink 2008):
\begin{equation}
F(r)=\int^{R}_{r} F_{d}(r^{\prime})\frac{r^{\prime}}{\sqrt{r^{\prime 2}-r^{2}}}dr^{\prime}
\end{equation}
where $R$ is the outermost radius of the object. The
deprojection was performed
from the outermost to the inner regions by using this formula.
In the deprojection process, the deprojected volume flux $F_{d}(r)$ is
 very sensitive to the fluctuations of
the observed surface flux $F(r)$. We therefore smoothed the $F(r)$ profile with
a Gaussian function of $FWHM=0.0235$ (or $1\sigma$ width of 0.01, with the outermost radius
scaled to 1), and then deprojected the smoothed one.
The inner boundary of the X-ray emitting ejecta shell was set to the position
where $F_{d}(r)$ first touches zero. Fig. 6 shows the comparison between the
observed radial profiles ($F(r)$) and those ($F_{p}(r)$) reprojected from $F_{d}(r)$, while
Fig. 7 shows the deprojected ones. As shown, $F_{p}(r)$ fits the observed surface
brightness $F(r)$ fairly well, especially in regions with $r>0.75$. However, in the inner
regions there exist some deviations. These deviations imply that $F(r)$ is not
the projection of a perfectly spherical emission distribution. Especially it
is relatively deficient of emission in the central region.
Therefore, the deprojected flux profiles are qualitatively reliable but
may not be correct quantitatively.

The errors of $F_{d}(r)$ were also obtained with a Monte-Carlo method.
For each distribution, using $F(r)$ as the expected
value and its error obtained in the spectral fitting as the 1-$\sigma$ width
of the Gaussian distribution, we simulated 100 surface flux distribution
curves, from which 100 deprojected volume flux curves have been
derived. Fig. 8 plots the distribution of these 100 simulated volume fluxes at each radius.
The uncertainty of the deprojected volume flux $F_{d}(r)$ plotted in Fig. 7
is actually the $1\sigma$ width of the distribution, similar to the
procedure shown in Fig. 4.

The continuum emission profiles in Figs. 6 (the 3rd row) and 7 are
inferred from the 0.85-1.5 keV
plus 3.0-10.0 keV fluxes with the nonthermal contribution subtracted. We select these
two energy bands so as to eliminate the influence of the strong Si and S emission lines.
The flux of the nonthermal emission is extrapolated by using the 4.0-6.0 keV
flux, because the 4.0-6.0 keV emission is dominated by the nonthermal
emission (e.g.,Warren et al. 2005; Eriksen et al. 2011; Lu et al. 2011).
In the flux extrapolation, the photon indices are not the same for
all the spectra. For those in regions 1 and 3 we use the photon index
of the nonthermal emission of the entire remnant, which is 2.7
(e.g., Warren et al. 2005;
Tamagawa et al. 2009). However, the spectral photon indices for the nonthermal
components in region 2 are slightly different. Eriksen et al.
(2011) found that the nonthermal strips in that region
have harder spectra than the nonthermal spectrum of
the entire remnant, the brightest
strip has an index of $\sim$2.1, and the other weaker strips have
a mean index of $\sim$2.5. Therefore, when we calculate the nonthermal
fluxes for the annuli in region 2, their photon indices are actually the
mean of those for the diffuse emission, the weak strips, and
the brightest strip weighted by their respective 4.0-6.0 keV photon
counts in each annulus, and the resulted photon indices are in the
range of 2.63-2.7. The 3rd to 5th rows of Fig. 6 plot the thermal
continuum emission, nonthermal emission, and the nonthermal to thermal
ratio profiles.

Now we estimate the uncertainties of the thermal emission flux that introduced
by the subtraction of the nonthermal component, which is in turn related to
the errors of the nonthermal flux. One error source of the nonthermal
flux is the contamination of the thermal emission. This contamination will
result in over-estimation of the nonthermal flux and then under-estimation of the
thermal flux. As shown in the bottom panel of
Fig. 6, except for the rims of the remnant, the nonthermal emission is only
about 5-20$\%$ as strong as the thermal emission. Considering that
the thermal contribution to the 4.0-6.0 keV energy flux is
$\sim$ 10$\%$ at the rims and increase to $\sim$ 40$\%$ in the inner
regions (Warren et al. 2005; Eriksen et al. 2011), we realise
that the thermal continuum flux could be over-subtracted up to $\sim 8\%$.
The other error source of the nonthermal emission is the variation of the
photon indices. As discussed above, the lowest photon index of the spectra
in region 2 could be 2.63 due to the existence of the harder
nonthermal strips, which implies that the photon indices used in this
paper might deviate from the actual ones up to 0.07.  Simple calculations
show that, if the 4.0-6.0 keV flux (in units of erg cm$^{-2}$ s$^{-1}$) is
the same, the 0.85-1.5 plus 3.0-10.0 keV flux of a spectrum
with photon index of 2.63 is
about 2.8$\%$ lower than that of a spectrum with photon index of 2.7. This
flux difference can be considered as the typical error of the nonthermal
fluxes caused by the variation of photon indices. As the 0.85-1.5 plus
3.0-10.0 keV thermal emission flux dominates the nonthermal flux except
at the rims (r$\geq$0.95), we conclude that the photon index variations of the
nonthermal emission have negligible influence on the determination of
the thermal fluxes. In summary, the typical errors of the thermal
continuum fluxes introduced by the subtraction of the nonthermal component
is less than $\sim 8\%$, much smaller than the overall variations of the
radial profiles.

The scaled electron density $n_{e}$ plotted in the 4th row of Fig. 4 is
simply the square root of the emission measure represented by the
deprojected thermal continuum fluxes (the 3rd row of Fig. 7).
Actually, the emission measure is proportional to
$n_e*n_H$, which is 1.18 $n_H^2$ (or 0.85 $n_e^2$) for plasma of
solar abundance, and  $n_e$ is proportional to the square root of
the thermal continuum fluxes with an accuracy of  $(1.18^{1/2}-1)=9\%$.
Decourchelle et al. (2001)  studied the spectra of several
typical regions in the Tycho's SNR. They found that in the knot where
the heavy element is most abundant, the Si abundance is 4.2 times of
solar value, S is 6.9, and Fe is 0.7. It is thus reasonable to
set the upper limit of the heavy element abundance as 7 times
solar values. If all these elements are fully ionized, $n_e$ is 1.24$n_H$, or the
emission measure is proportional to 0.80 $n_e^2$. In this case, $n_e$
is proportional to the square root of the thermal continuum fluxes
with an accuracy better than  $(1.24^{1/2}-1)=11\%$.
We also note that Hwang et al. (1998) obtained a much higher abundance
of about 30 times solar value for heavy elements such as Si and S.
Even if all the elements are that abundant and fully ionized,
$n_e$ will be 1.48$n_H$, and $n_e$ will be proportional to the square
root of the thermal continuum fluxes with an accuracy about $(1.48^{1/2}-1)=22\%$.
Based on these discussions we concluded that the elemental
abundance variation in the ejecta can not change much the radial profiles
of $n_e$ shown in the 4th row of Fig. 7.

As can be seen in Fig. 7, the S/Si line ratios increase almost
monotonically with $r$, while $n_e$ have big bumps at $r$ between
0.9 and 0.75, which means that $t$ must increase with radius and
dominate the $\tau$ variations. So the ejecta in the outer region
were heated earlier. This is just what the RS scenario means.
 For reference, we inferred from the S/Si line ratio the $\tau$
values under the assumption that the temperature is constant
(1.0 keV) and the S/Si abundance ratio is 1.35 solar
value (Hwang \& Gotthelf 1997). As plotted  in the bottom
panel of Fig. 7, the decrease of $\tau$ to the remnant center
is quite obvious.

For the sake of the argument, we outline the consequences for the case that
the plasma emitting the bulk of the Si and S lines were heated by a forward shock.
The pre-shock medium were required to be enriched in Si and S and should
have a density
that increases with radius. Such a profile is very unlikely to exist, neither
for the ejecta shortly after the SN explosion nor for a putative
ambient interstellar
medium. The stellar matter ejected a long time before the explosion
could have such a density distribution. But there is no observational evidence
at any other wavelength for the existence of such a shell or ring,
strongly supporting
our findings that $t$ increases with radius.

\section{Summary}

Using the long exposure {\it Chandra} ACIS-I observations of Tycho's SNR, we
have obtained radial distributions of the He-like K$\alpha$ lines of Si and S
for three fan-like regions where the brightness distributions look
spherically symmetric. In all
the three regions, both the observed and the deprojected S/Si line flux ratios
increase almost monotonically with distance from the remnant center. We
have shown that this line flux ratio is a good tool to investigate the shock wave
kinematics in young SNRs, and the observed results can only be explained if the
ejecta in the outer regions were heated earlier. Therefore, we provide for the
first time radial profiles of the RS action in Tycho's SNR, equivalent to the
temporal evolution of the RS. Furthermore, our study may
also provide a potential new method to derive the time development of the RS
velocity giving additional insight to the hydrodynamic processes in young SNRs
 by eventually more detailed
studies of radial variations of the S/Si line flux ratios as well as
the plasma temperatures
and densities.

\section*{Acknowledgments}

F.J.L and M.Y.G contributes equally to this work. We thank
Q.Daniel Wang, Yang Chen, Aigen Li and Mei Wu for very useful suggestions. F.J.L
thanks Kenichi Nomoto for helpful discussions. This work is supported by grants from
National Science Foundation of China (11233001,11133002).

\clearpage

\begin{deluxetable}{ccc}
\tabletypesize{\footnotesize}
\tablecaption{ Chandra ACIS-I observations of Tycho's SNR in 2009. \label{obsid_table}}
\tablewidth{0pt}
\tablehead{\colhead{ObsID}   & \colhead{Start Date}   &\colhead{Exposure(\,ks)}}
\startdata
10093& 2009-04-13 & 118    \\
10094& 2009-04-18 & 90     \\
10095& 2009-04-23 & 173    \\
10096& 2009-04-27 & 106    \\
10097& 2009-04-11 & 107    \\
10902& 2009-04-15 & 40     \\
10903& 2009-04-17 & 24     \\
10904& 2009-04-13 & 35     \\
10906& 2009-05-03 & 41     \\
Total&            & 734.11 \\
\enddata
\end{deluxetable}

%\begin{deluxetable}{cccc}
%\tablecaption{ 3 region information of Tycho's supernova remnant. \label{reg_table}}
%\tablewidth{0pt}
%\tablehead{\colhead{Reg}   & \colhead{Start Degree($^{\circ}$)} & \colhead{Stop Degree($^{\circ})$}
%&\colhead{Radius($^{\prime\prime}$)}}
%\startdata
%1 & 55  &68  & 228    \\
%2 & 310 &390 & 258    \\
%3 & 90  &130 & 252    \\
%\enddata
%\end{deluxetable}

%%%%%%%%%%%%%%%%%%%%%%%%%%%%%%%%%%%%%%%%%

\begin{deluxetable}{cclll}
\tabletypesize{\footnotesize}
\tablecaption{The fitted parameters of the 8 Gaussian components
representing the Si and S line emission. \label{Line_talble}}
\tablewidth{0pt}
\tablehead{\colhead{Elements} & \colhead{Components}
& \colhead{Central Energy(\,keV)}   & \colhead{Width(\,keV)}
& \colhead{Flux($10^{-15}$\,ergs\,cm$^{-2}$\,s$^{-1}$)}}

\startdata
    & G1 & 1.853 (1.852, 1.854) & 0.028 (0.026, 0.030)  & 10.29 (10.06, 10.43) \\
 Si & G2 & 2.006$^{\ast}$   & 1E-5$^{\ast}$  & 0.83(0.74, 0.92)  \\
    & G3 & 2.182  & 1E-5 & 1.11 (1.05, 1.16) \\
    & G4 & 2.294  & 1E-5 & 0.61 (0.58, 0.64) \\\hline
    & G1 & 2.451 (2.448, 2.454) & 0.019 (0.003, 0.027) & 3.16 (3.06, 3.27) \\
  S & G2 & 2.623 & 1E-5  & 0.12 (0.05, 0.18) \\
    & G3 & 2.884 & 1E-5  & 0.22 (0.20, 0.25) \\
    & G4 & 3.033 & 1E-5  & 0.12 (0.11, 0.14)
\enddata
\\
$\ast$no error range given means that the parameter is fixed.
\end{deluxetable}

%%%%%%%%%%%%%%%%%%%%%%%%%%%%%%%%%%%%%%%%%%%%%%%%%%%%%%%%%%%%%%%%%%%%%%%%%%%%%%%%
%%%%%%%%%%%%%%%%%%%%%%%%         Figure       %%%%%%%%%%%%%%%%%%%%%%%%%%%%%%%%%%
%%%%%%%%%%%%%%%%%%%%%%%%%%%%%%%%%%%%%%%%%%%%%%%%%%%%%%%%%%%%%%%%%%%%%%%%%%%%%%%%

\clearpage
\begin{figure}
\includegraphics[width=1.0\textwidth,angle=0]{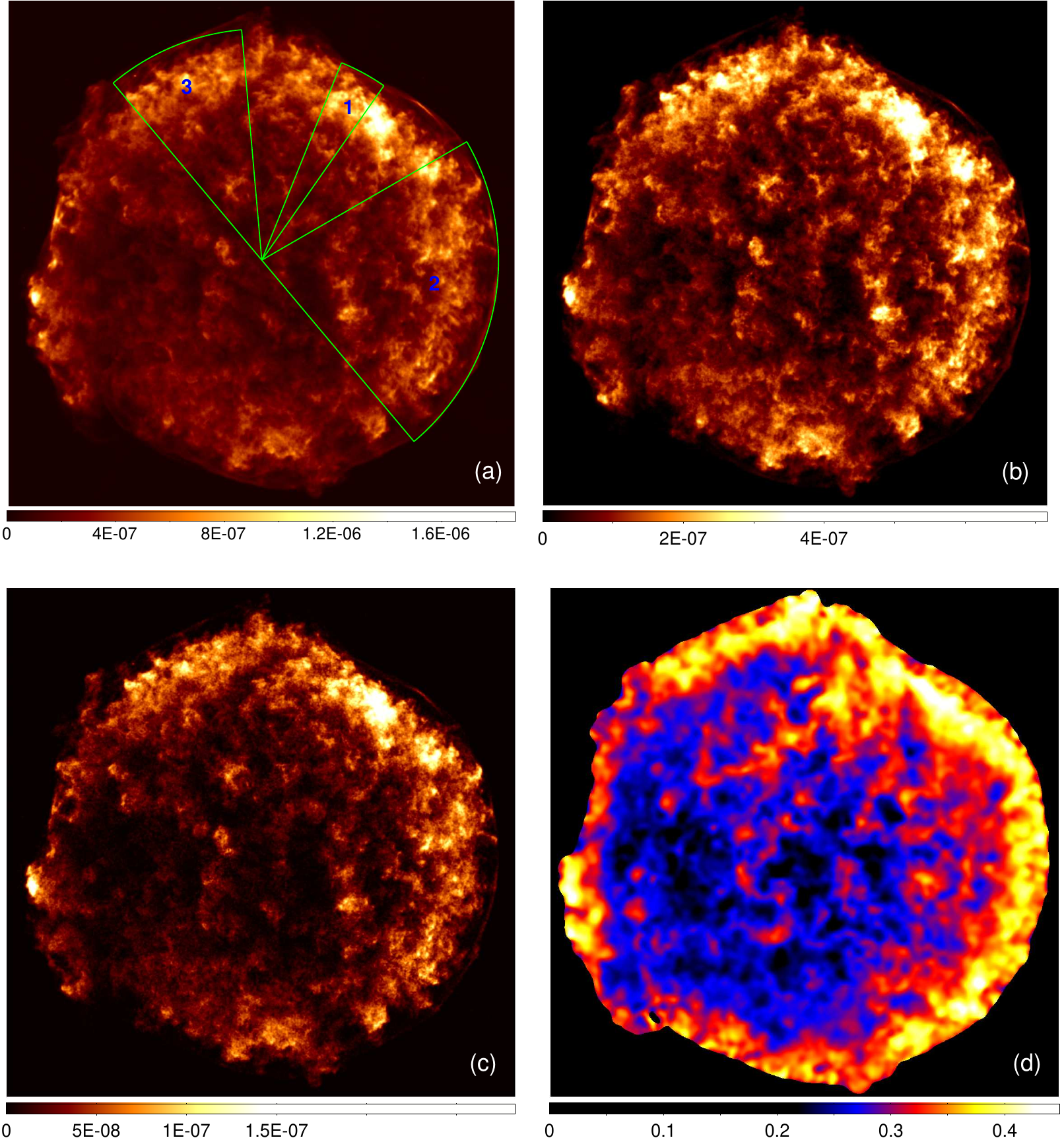}
\caption{The X-ray intensity maps for the Tycho's SNR in
0.5-8.0\,keV(a), 1.6-2.0\,keV(b)
and 2.2-2.6\,keV(c), as well as the 2.2-2.6 to 1.6-2.0\,
keV intensity ratio map (d) for this remnant.
The images are obtained with $Chandra$/ACIS-I, and the
intensities are in units of cts arcsec$^{-2}$ s$^{-1}$.
(b) is dominated by Si K$\alpha$ line, and (c) by the
S K$\alpha$ line. (d) has been Gaussian-smoothed with the
FWHM of 10$\arcsec$, and the ratio of low brightness region
outside the remnant has been set as zero.
The three fan-like regions in (a) denote those for which
spectral analyses have been performed.
\label{Fig1}}
\end{figure}

\clearpage

\begin{figure}
\includegraphics[width=0.8\textwidth,angle=0]{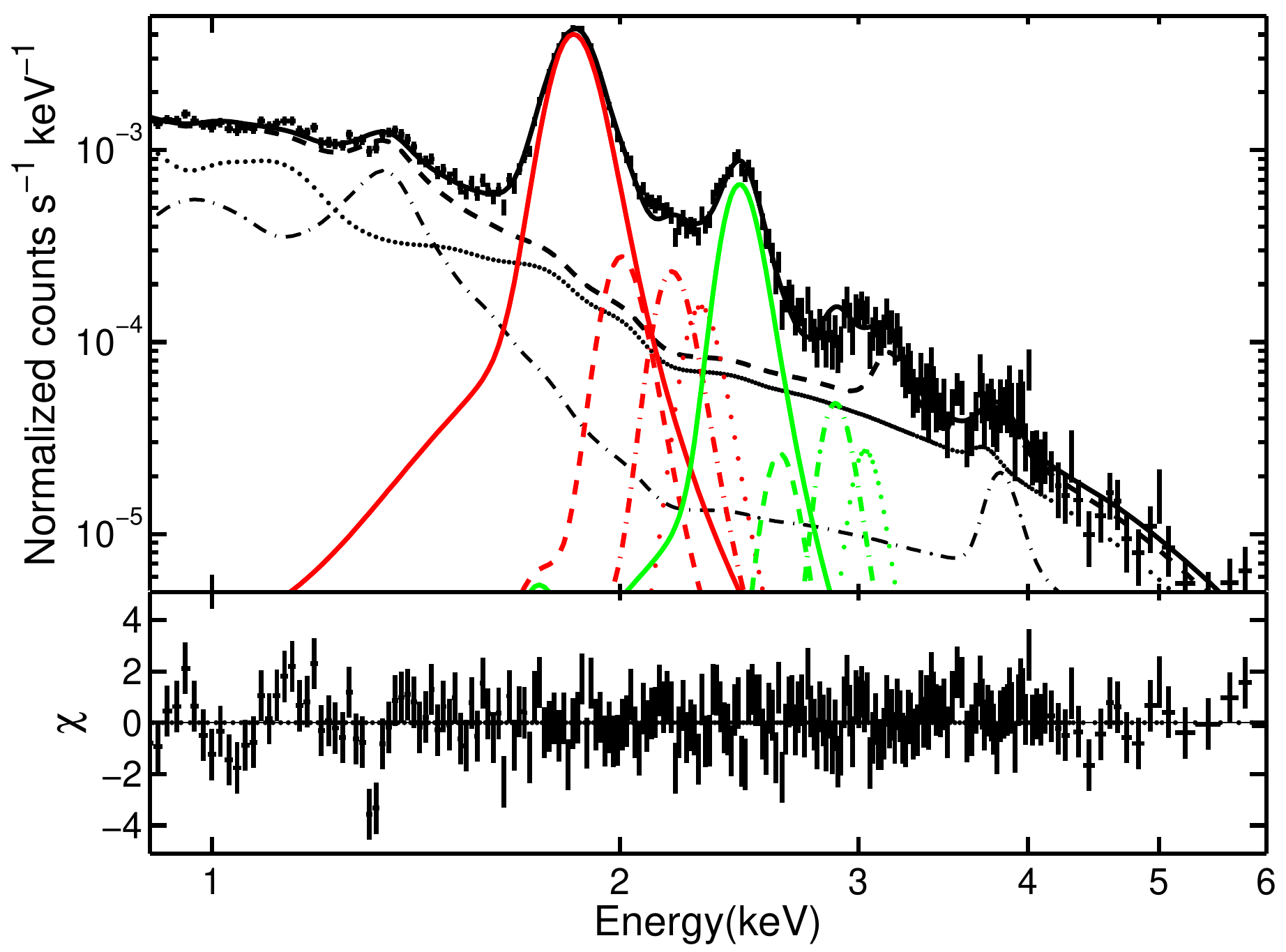}
\caption{The X-ray spectrum of an annulus in region 3 with
radius of 0$^{\prime\prime}$ to 24$^{\prime\prime}$.
Upper panel: The solid black line represents the two-component
VNEI model fitted to the data,
and the dashed black line represents the same model but setting
the Si and S abundances to zero.
The red lines are the four components used to account for the
Si line emission, while the four
green lines are for S. The dotted and dash-dotted black lines
are the two VNEI components
respectively with the Si and S abundances fixed at zero.
Lower panel: the residual of the
(2 VNEI + 8 Gaussian) model fit to the data. The Si and S
abundances in the 2 VNEI model have
been set to zero as mentioned above.
\label{Fig3}}
\end{figure}

\clearpage

\begin{figure}
\includegraphics[width=1\textwidth,angle=0]{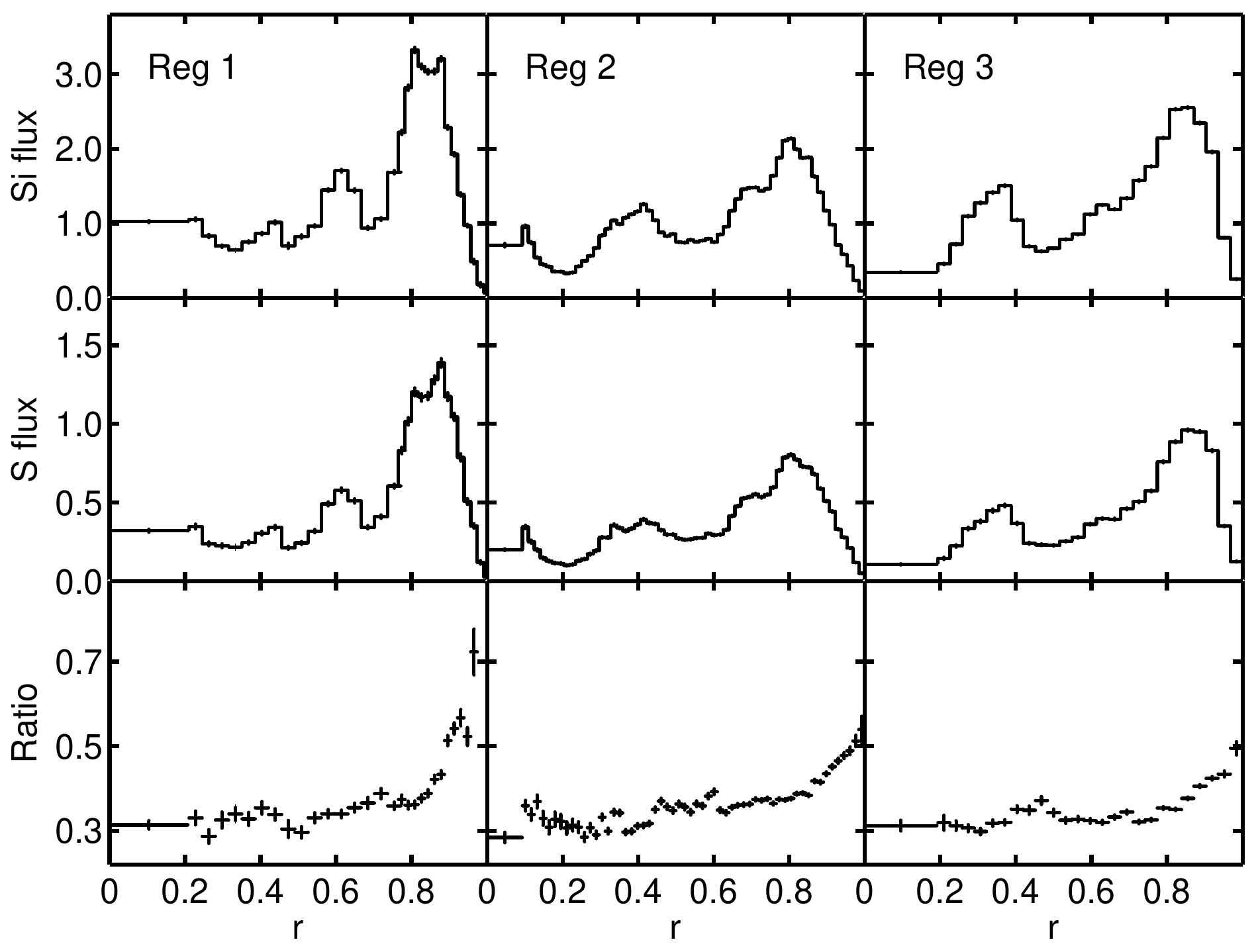}
\caption{The Si and S He-like K$\alpha$ line flux properties as
a function of radius ($r$) from the remnant center
(RA(2000)=00$^{h}$ 25$^{m}$ 19.4$^{s}$,
DEC(2000)=64° 08$^{\prime}$ 13.98$^{\prime\prime}$)
for the three regions denoted in Fig. 1a. The
fluxes are in units of 10$^{-15}$\,ergs\,cm$^{-2}$\,s$^{-1}$\,arcsec$^{-2}$,
and the ratio is S to Si.
$r$ is scaled to that of the blast wave. The bump
at $r\sim0.45$ in the bottom right plot is due to a bright
filamentary fragment as can be seen in region
3 of Fig. 1(a). The upper-left (outer) edge of this fragment
is heated earlier and so shows S/Si line
ratio profile resembling that of the whole section.
The error bars show the $1\sigma$ uncertainties.
\label{Fig2}}
\end{figure}

\begin{figure}
\includegraphics[width=0.8\textwidth,angle=0]{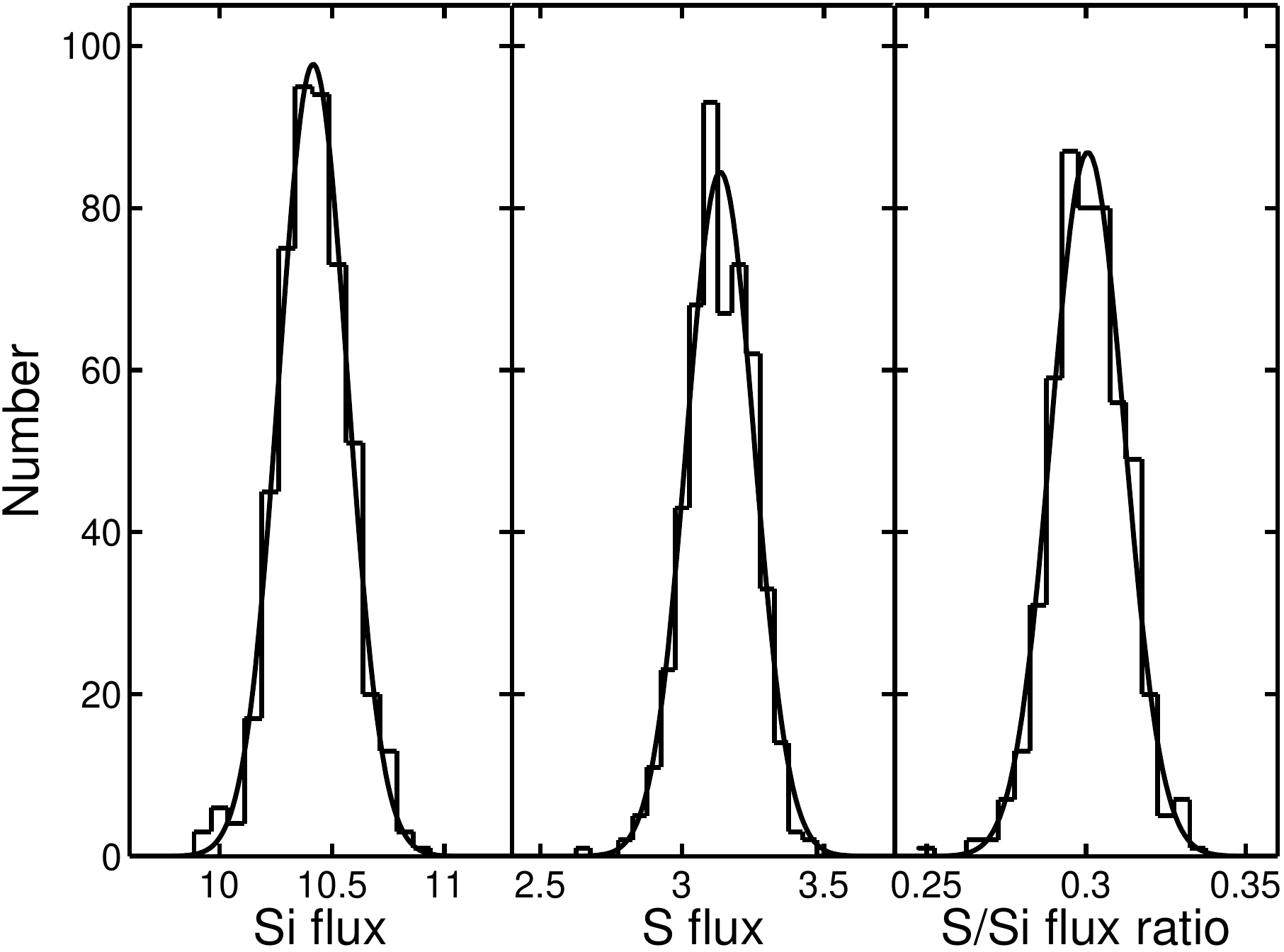}
\caption{The distribution of the 500 simulated Si, S line fluxes
and S/Si line flux ratios.
It shows that the $1\sigma$ statistical errors of Si flux,
S flux and S/Si flux ratio are
about 1.5$\%$, 3.7$\%$, and 3.9$\%$. The units of the fluxes
are $10^{-15}$ ergs cm$^{-2}$ s$^{-1}$.
\label{Fig4}}
\end{figure}

\clearpage

\begin{figure}
\includegraphics[width=1\textwidth,angle=0]{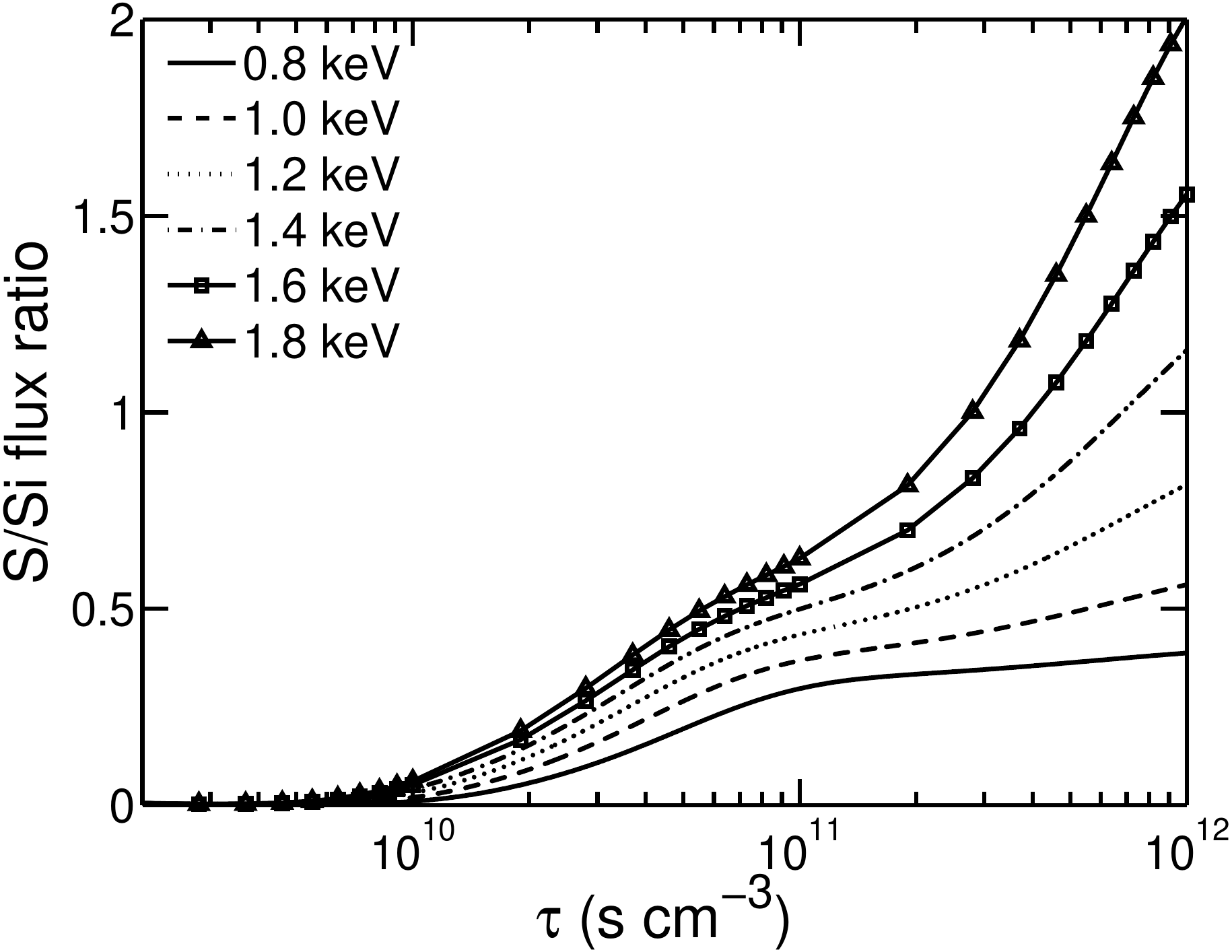}
\caption{The relation between the He-like S to Si K$\alpha$ line
ratio and ionization age $\tau$ for different
plasma temperatures. The results are calculated with the VNEI
model in XSPEC, and the S/Si abundance
ratio is fixed at 1.35 times solar value as obtained by
Hwang \& Gotthelf (1997) for the Tycho's SNR.
\label{Fig5}}
\end{figure}

\begin{figure}
\includegraphics[width=1\textwidth,angle=0]{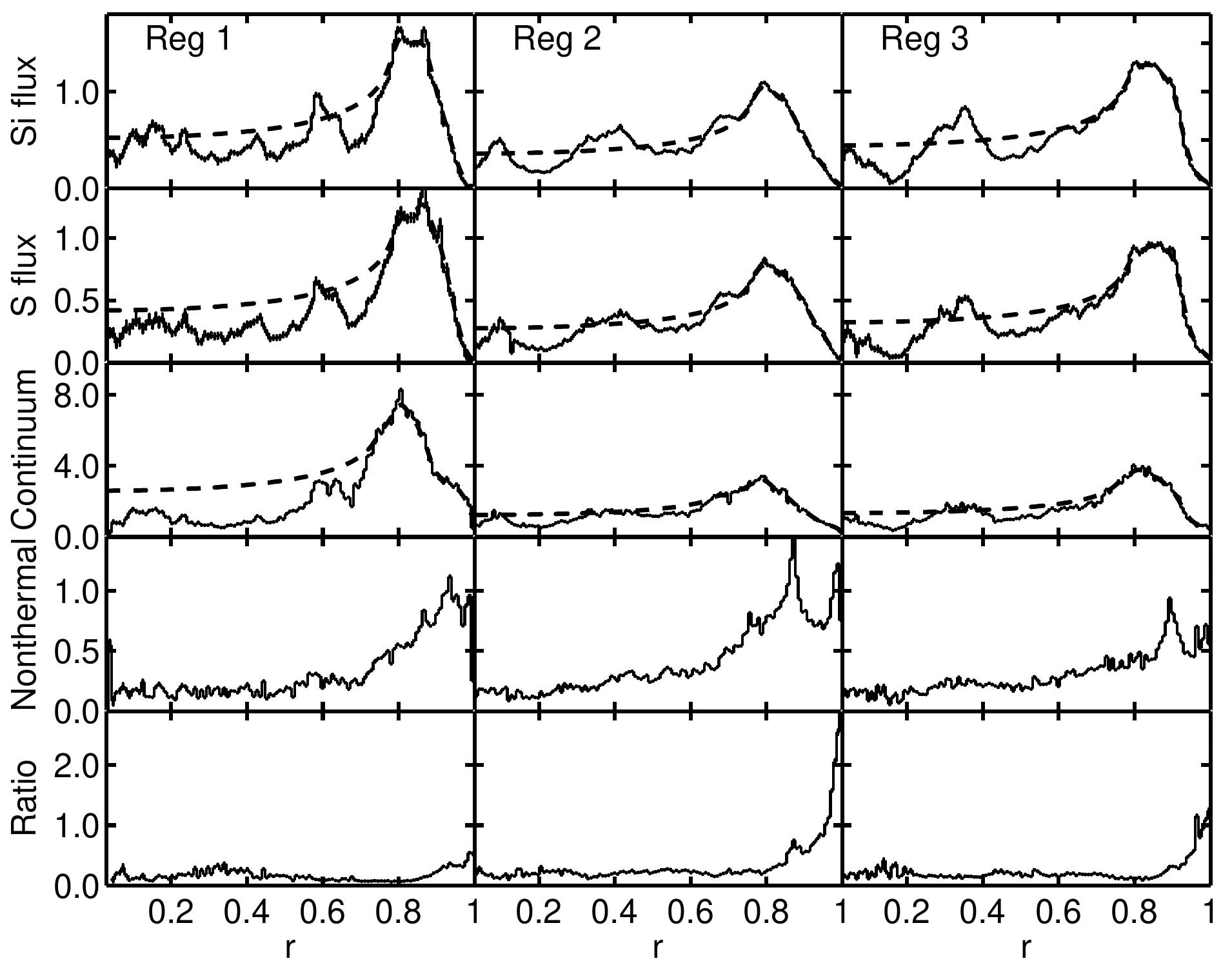}
\caption{The measured radial surface flux distribution (solid lines)
of the Si (top panel) line,
the S (the 2nd panel) lines, the thermal continuum emission
(0.85-1.5\,keV plus 3.0-10.0\,keV;
the 3rd panel), the nonthermal emission (0.85-1.5\,keV plus 3.0-10.0\,keV;
the 4th panel), the nonthermal to thermal continuum emission ratio (bottom panel),
as well as the reprojected Si, S, and thermal continuum
profiles (dashed lines). The fluxes are in units of
10$^{-15}\,$ergs\,cm$^{-2}$\,s$^{-1}$\,arcsec$^{-2}$, and the
horizontal axis ($r$) has the same meaning as in Fig. 3.
\label{Fig6}}
\end{figure}

\clearpage

\begin{figure}
\includegraphics[width=1\textwidth,angle=0]{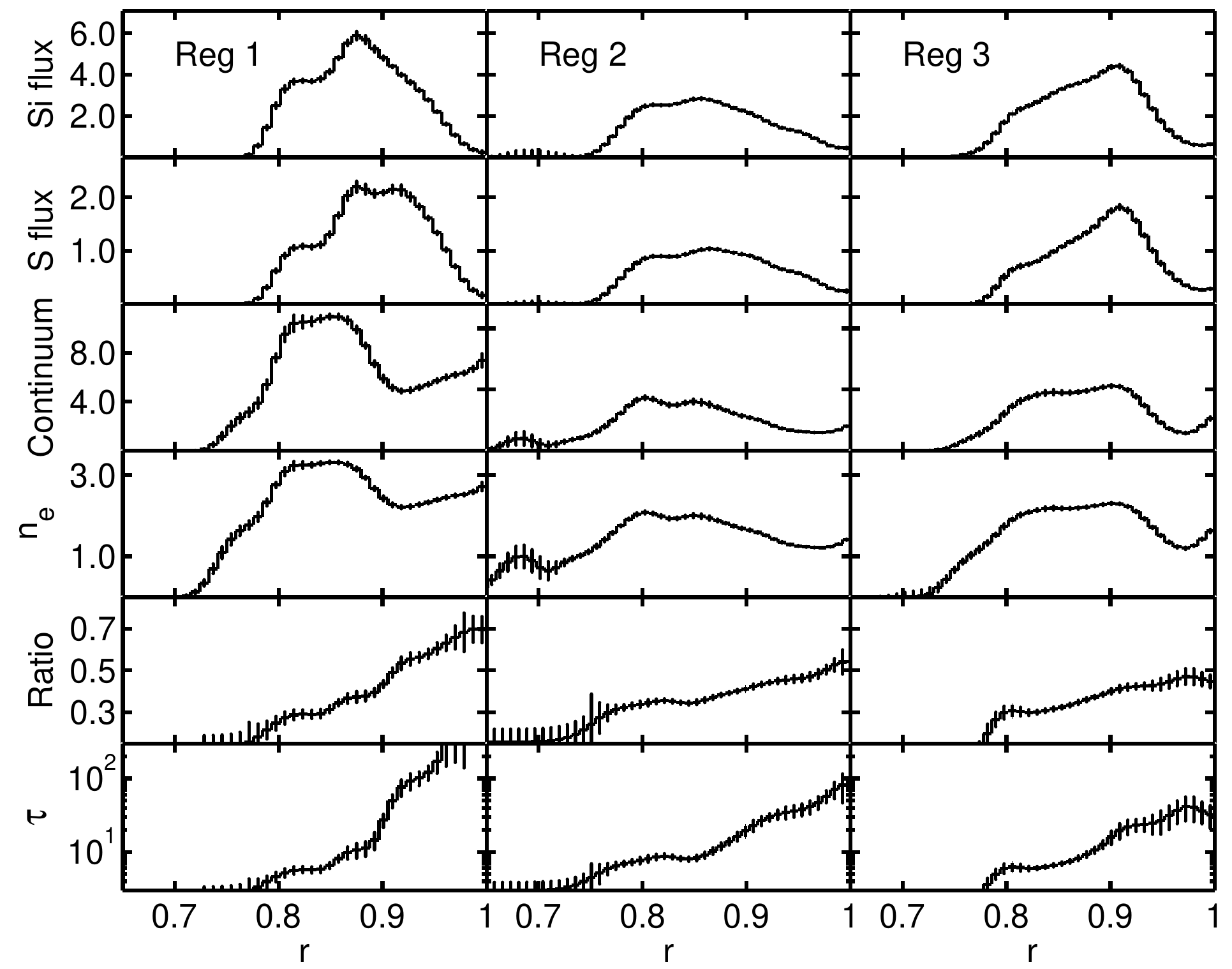}
\caption{The radial profiles of the deprojected volume fluxes of the Si, S,
and thermal continuum emission, as well as the scaled electron number
density $n_e$, S/Si flux ratio, and the ionization age $\tau$ inferred
from the line flux ratio assuming a constant temperature of 1.0 keV and
S/Si abundance ratio of 1.35 times the solar value (Hwang \& Gotthelf 1997).
The fluxes are in units
of 10$^{-18}$\,ergs\,cm$^{-2}$\,s$^{-1}$\,arcsec$^{-3}$.
The continuum emission fluxes are for 0.85-1.5\,keV plus 3.0-10.0\,keV band with
the nonthermal component subtracted, and the scaled electron number density $n_e$
is simply a square root of the volume flux of the thermal continuum emission,
because the flux is roughly proportional to $n_{e}^{2}$ as discussed in
the text. The units of the scaled $n_e$ are artificial, and units of $\tau$ are
$10^{10}$ cm$^{-3}$ s. In the deprojection,
a spherical symmetry is assumed. The error bars are obtained with a
Monte-Carlo method as described in the text and Fig. 8. $r$ has the
same meaning as in Fig. 3.
\label{Fig7}}
\end{figure}

\begin{figure}
\includegraphics[width=0.8\textwidth,angle=0]{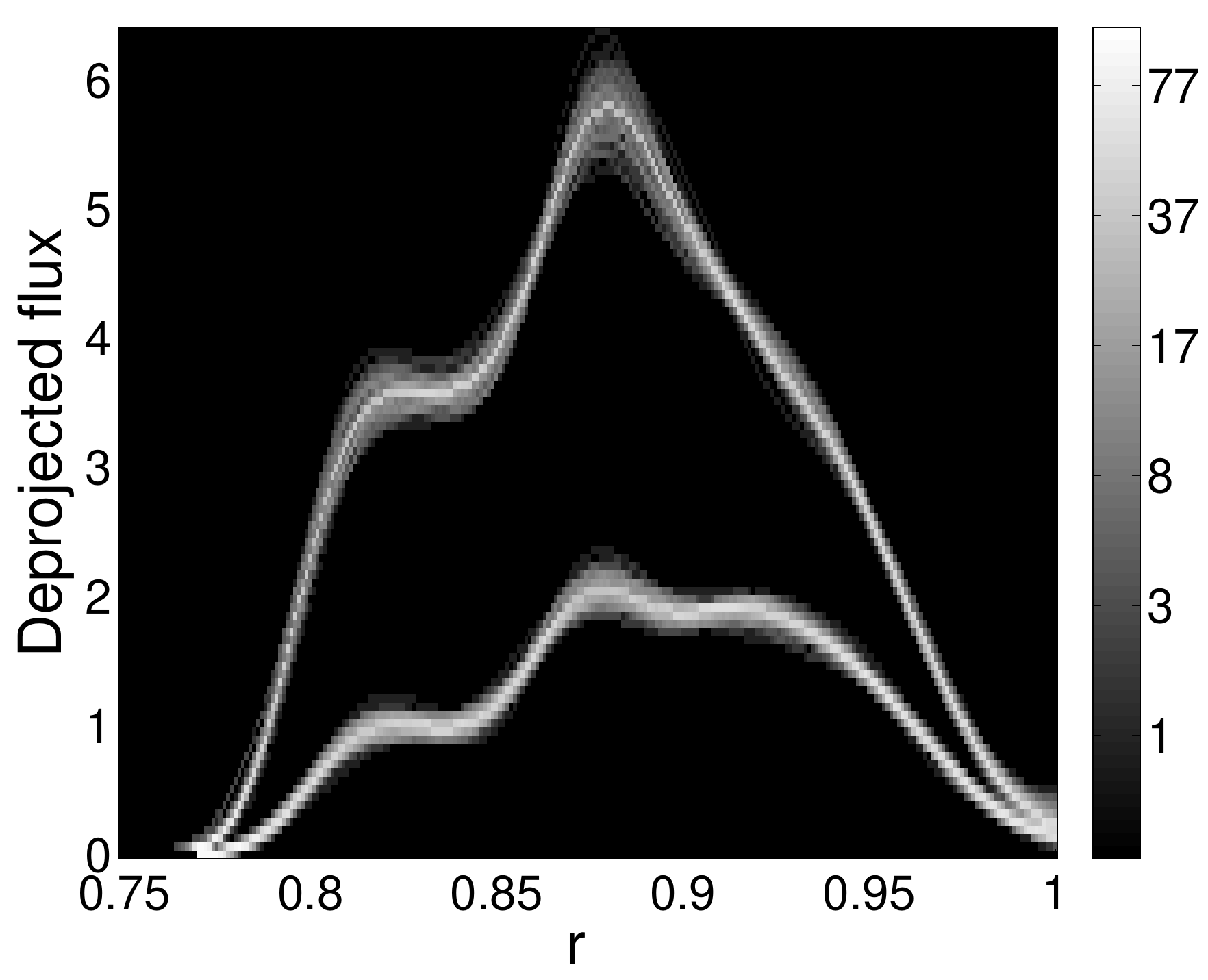}
\caption{Distributions of the deprojected volume fluxes of the Si (the upper belt)
and S (the lower belt) lines from the simulated surface brightness profiles
for region 1 that defined in Fig. 1(a). The brightness of the belts represents the
number of the simulated volume fluxes in the range defined by the vertical axis
(in units of $10^{-18}$\,ergs\,cm$^{-2}$\,s$^{-1}$\,arcsec$^{-3}$). $r$ has the same
meaning as in Fig. 3.
\label{Fig8}}
\end{figure}

\clearpage

\end{document}